\documentclass[]{jfm}
\usepackage{graphicx}
\usepackage{newtxtext}
\usepackage{newtxmath}
\usepackage{natbib}
\usepackage{amsmath}
\usepackage{hyperref}
\hypersetup{
    colorlinks = true,
    urlcolor   = blue,
    citecolor  = black,
}

\newcommand{\RomanNumeralCaps}[1]
\linenumbers

 \usepackage{algorithmic}
\usepackage{algorithm2e}

\usepackage{listings}

\title{Casting Computational Fluid Mechanics into a Convex Quadratic Optimization Framework}

\author{Hussam Sababha\aff{1}
  \corresp{\email{haa385@nyu.edu}}, Haithem Taha\aff{2} \and Mohammed Daqaq\aff{1}}
\affiliation{\aff{1}Division of Engineering, New York University Abu Dhabi, Abu Dhabi, UAE\aff{2}University of California Irvine, Samueli School of Engineering, CA, USA}

\begin{document}
\maketitle

\begin{abstract}
 We employ the principle of minimum pressure gradient to transform problems in unsteady computational fluid dynamics (CFD) into a convex optimization framework subject to linear constraints. This formulation permits solving, for the first time, CFD problems efficiently using well-established quadratic programming tools or using the well-known Karush-Kuhn-Tucker (KKT) condition. The proposed approach is demonstrated using three benchmark examples. In particular, it is shown through comparison with traditional CFD tools that the proposed framework is capable of predicting the flow field in a lid-driven cavity, in a uniform pipe (Poiseuille flow), and that past a backward facing step. The results highlight the potential of the method as a simple, robust, and potentially transformative alternative to traditional CFD approaches.
\end{abstract}


\section{Introduction}
\label{sec:intro}
Although variational methods have demonstrated success in many fields of mechanics \citep{lanczos2012variational}, their implementation in fluid mechanics remains relatively limited \citep{Variational_Principles_Fluids_HamiltoN_poF}. This stems from the difficulty of incorporating dissipative force such as viscosity using the standard Hamiltonian formulation \citep{bretherton1970note, salmon1988hamiltonian, morrison1998hamiltonian, Morrison2020lagrangian}. As a result, traditional computational methods in fluid mechanics remain predominantly confined to Newtonian mechanics where the primary focus is on the solution of the Navier-Stokes equation (NSE) or one of its derivatives \citep{gresho1987pressure}.

In a departure from traditional variational principles that are based on least action, Taha et al. \citep{gonzalez2022variational, taha2023minimization} used Gauss' principle of least constraint \citep{Gauss_Least_Constraint,Papastavridis,Udwadia_Kalaba_Original,udwadia2002foundations,Udwadia_Kalaba_Book,udwadia2023general} to develop a new variational principle in fluid mechanics known as the principle of minimum pressure gradient (PMPG). This principle asserts that an incompressible flow evolves from one instant in time to another such that the total magnitude of the pressure gradient over the domain is minimized. Taha et al. \citep{taha2023minimization} proved that the necessary condition to minimize the cost of the pressure gradient is guaranteed to satisfy the NSE, which transforms a fluid mechanics problem into an optimization problem. Taha et al. \citep{taha2023minimization} employed PMPG to find analytical solutions to some classical problems in fluid mechanics, including unsteady laminar flow in a channel and Stokes' second problem. The attained solutions provided additional insights that are otherwise difficult to infer using traditional approaches \citep{gonzalez2022variational, shorbagy2024magnus}. 

The development of computational frameworks based on PMPG remains limited and is still in its early stages. Recent work has utilized Physics-Informed Neural Networks (PINNs) to solve the optimization problem resulting from PMPG \citep{sababha2024variational, alhussein2024principle, atallah2024novel}. The proposed method, validated against benchmark examples, demonstrated two key advantages. First, it eliminates the pressure-velocity coupling, thereby alleviating the computationally expensive step of solving the Poisson equation at every time step, which often dominates the runtime in traditional algorithms. Second, it removes the need to impose nonphysical outflow boundary conditions, allowing for a reduction in the computational domain required for most fluid mechanics problems. Despite their advantages, the use of PINN for optimization presents several challenges. First, accurately capturing the complexity of fluid flows often necessitates large neural networks, significantly increasing the number of trainable parameters, such as weights and biases. Second, constraints are typically enforced in a soft manner, making it challenging to guarantee their satisfaction throughout the entire domain. Finally, the optimization process remains inherently non-convex, which can result in multiple, potentially inconsistent solutions, as highlighted in previous studies \citep{he2023artificial, wang2023solution}.

In this work, we transform the PMPG computational framework into a quadratic optimization framework. By focusing on the local acceleration as the primary variable, we reformulate the minimization problem into a convex structure subject to linear constraints. This permits the implementation of well-established quadratic programming tools for optimization or a direct solution through the well-known Karush-Kuhn-Tucker (KKT) condition. We demonstrate the effectiveness of this approach and highlight its simplicity using three benchmark examples; namely the flow in a led-driven cavity, the Poiseuille flow, and the flow past a backward step. 

The remainder of this paper is organized as follows. Section 2 provides a review of the theoretical foundation of the PMPG. Section 3 presents the formulation of the problem. Section 4 presents the results. Section 5 presents concluding remarks that discuss the advantages and limitations of the current approach, as well as potential extensions.

\section {Fluid Mechanics as a Convex Minimization Problem}
In the context of the constrained motion of particles, Gauss' principle of least constraint states that the constrained system accelerates so that it minimizes, in a weighted least-squares sense, the deviation between its current acceleration and that of the free motion \citep{Gauss_Least_Constraint}. Thus, a particle adjusts its path only to the extent necessary to meet the constraints, ensuring the least possible deviation from the unconstrained trajectory; i.e., its free/natural motion.  

In mathematical terms, for $N$-constrained particles, each of mass, $m_i$, whose motion is described by $\textbf{q}$ generalized coordinates, Gauss' principle asserts that the quantity 
\begin{equation}
    \mathcal{A} = \frac{1}{2} \sum_{i=1}^{N} m_i \left( \mathbf{a}_i (\mathbf{q}, \dot{\mathbf{q}}, \ddot{\mathbf{q}})- \frac{\mathbf{F}_i}{m_i}\right)^2,
\end{equation}
also known as the Gaussian cost, is a minimum with respect to the generalized accelerations, $\ddot{\mathbf{q}}$, at every instant of time. Here, $\mathbf{a}_i$ and $\mathbf{F}_i$ are, respectively, the acceleration of the $i^{th}$ particle, and the net (non-constraint) force acting on it. Gauss' principle, therefore, turns the mechanics problem governed by Newton's Equation into the instantaneous minimization problem.

Using the NSE and the definition of the Gaussian cost,  Taha et al. \citep{taha2023minimization} extended Gauss' principle of least constraint to a continuum of fluid particles forming an incompressible fluid. They showed that, in the case of some fluid domain, $\Omega$,  the Gaussian cost can be written in Eulerian coordinates as:
\begin{equation}
    \mathcal{A} = \displaystyle \frac{1}{2} \rho  \int_{\Omega} \left(\frac{\partial \mathbf{u}}{\partial t} + \mathbf{u} \cdot \nabla \mathbf{u} - \nu \Delta \mathbf{u} \right)^2 d\mathbf{x}, \label{actionS}
\end{equation}
where, $\rho$, is the density of the fluid, $\mathbf{u}$ is the fluid velocity vector,  and $\nu$ is the kinematic viscosity. 

Taha et al. \citep{taha2023minimization} further demonstrated that the solution, $\textbf{u}$, which satisfies NSE and the continuity, $\nabla . \mathbf{u}= 0$, subject to any boundary conditions, is the same as the one that minimizes the functional, $\mathcal{A}$, in Equation (\ref{actionS}) with respect to the local acceleration, $\frac{\partial \mathbf{u}}{\partial t}\equiv \mathbf{u}_t$ subject to $\nabla . \mathbf{u}_t = 0$ and the same boundary conditions. Thus, for an incompressible fluid domain, $\Omega$, bounded by the surface, $\Gamma$, the problem of finding $\mathbf{u}$, which satisfies continuity and NSE can be cast in the following form:

Find $\mathbf{u}_t$ such that

\begin{alignat}{3}
\min_{\textbf{u}_t} &\quad&  \mathcal{A}(\textbf{u}_t) & \qquad  \textbf{x} \in \Omega,&  \\
\text{subject to: } &\quad&  \nabla . \mathbf{u}_t = 0 &\qquad \textbf{x} \in \Omega,&\\  
                    &\quad&  \mathbf{u}_t= 0     & \qquad   \textbf{x} \in \Gamma.&        
\end{alignat}

A fundamental advantage of this formulation is that the expression in Equation \eqref{actionS} is quadratic in $\mathbf{u}_t$, and the constraints are linear. Thus, the solution scheme lends itself naturally to the quadratic optimization family whose solution can be obtained using standard optimization techniques that are straightforward to implement and computationally efficient.

\section{Problem Formulation}
To solve the minimization problem, we consider a structured domain with equispaced collocation points arranged in an \( n_x \times n_y \) grid. At each node, the optimization variables are $\mathbf{u_t} = (u_t, v_t)$, representing the local accelerations in the $x$ and $y$ directions, respectively. Thus, the total number of optimization variables is $2n_xn_y$. The optimization vector, here denoted as \( \mathbf{w} \), is constructed by first flattening the $u_t$ and $v_t$ components from the grid, and then concatenating them; transforming the representation from a structured grid into a single optimization vector as:
\begin{equation}
\mathbf{w} = 
\begin{bmatrix}
u_{t,1}, u_{t,2}, \dots, u_{t,n^2}, v_{t,1}, v_{t,2}, \dots, v_{t,n^2}
\end{bmatrix}^T.
\end{equation}

Using the mean rule and the definition of the vector \( \mathbf{w} \), the integral of the objective function $\mathcal{A}$ can be re-written in a quadratic form as
\begin{equation}
    \mathcal{A} = \frac{1}{2} \mathbf{w}^T \mathbf{H} \mathbf{w} + \mathbf{f}^T \mathbf{w},
    \label{eq:obj}
\end{equation}
where $H$ is an $2n_xn_y \times 2n_xn_y$ positive definite matrix often referred to as the Hessian matrix, and $f$ is a $2n_xn_y \times 1$ vector representing external forces arising from the convective and viscous terms. It is worth noting that the objective function in this form excludes constant terms, as they do not influence the optimization process.

Next, we rewrite the continuity constraint (Equation (2.4)) in terms of the vector $\mathbf{w}$ using the finite difference method. In particular, we approximate the continuity equation at an interior point $(x_i, y_i)$ by using the forward difference formula:
\begin{equation}
 \frac{u_t(x_{i+1}, y_i) - u_t(x_{i}, y_i)}{h} + \frac{v_t(x_{i}, y_{i+1}) - v_t(x_{i}, y_{i})}{h} = 0  , \quad \text{for } i,j = 1, \ldots, n-1.
 \label{forward}
\end{equation}
Using Equation (\ref{forward}), the continuity equation can be expressed as $D \textbf{w}=0$, where the elements of the matrix \( D \) at each interior point \( i, j \) have non-zero entries at \( D_{i,i-1} = -\frac{1}{h} \), \( D_{i,j-1} = \frac{1}{h} \)  and \( D_{i,j} = \frac{1}{h} \). The rest of the elements are zero making the system extremely sparse.

Finally, the boundary conditions are represented by the condition $B \textbf{w}=0$, where the matrix $B$ is chosen to ensure that $\mathbf{w}$ vanishes along the four boundaries. By combining the continuity and boundary constraint matrices, the augmented system of constraints becomes:
\begin{equation}
A \mathbf{w} = 
\begin{bmatrix}
D \\
B
\end{bmatrix} \mathbf{w} = 
\begin{bmatrix}
\mathbf{0} \\
\mathbf{0}
\end{bmatrix}.
\label{eq:bc}
\end{equation}

\begin{figure}
\centering
\includegraphics[width=0.55\textwidth]{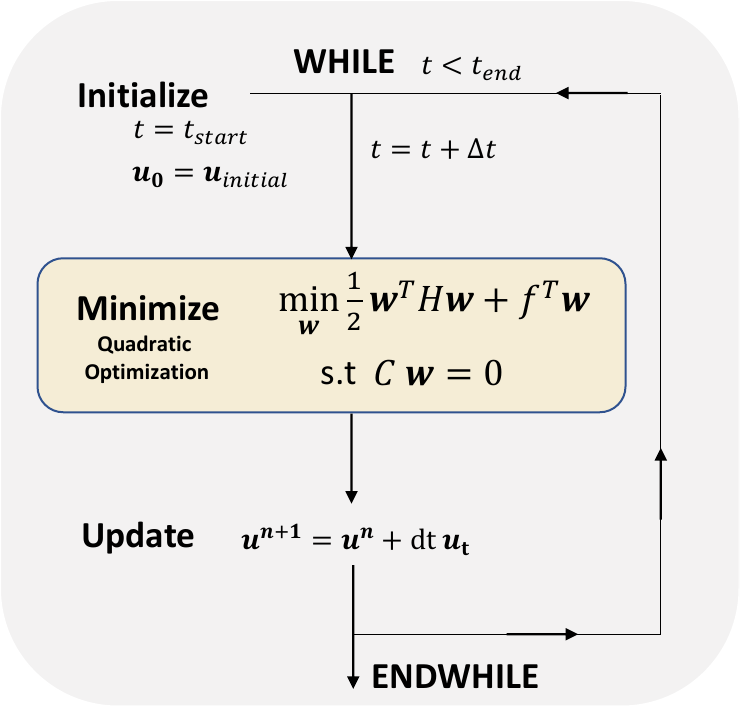}
\caption{A flow chart illustrating the solution methodology using the proposed approach.}\label{flowchart1}
\end{figure}

Equations \eqref{eq:obj} and \eqref{eq:bc} represent the standard form of a quadratic optimization problem, which can be solved in a finite amount of computational time \citep{nocedal2006quadratic}. The computational effort required to find a solution is influenced by the characteristics of the objective function. In this case, the matrix $H$ is positive definite, making the problem strictly convex. Strictly convex problems are computationally similar to linear programming problems, which are well-known for their efficient solution methods \citep{nocedal2006penalty}.The solution methodology is outlined in the flowchart~\ref{flowchart1}. 

\section{Results}
\begin{figure}
\centering
\includegraphics[width=\textwidth]{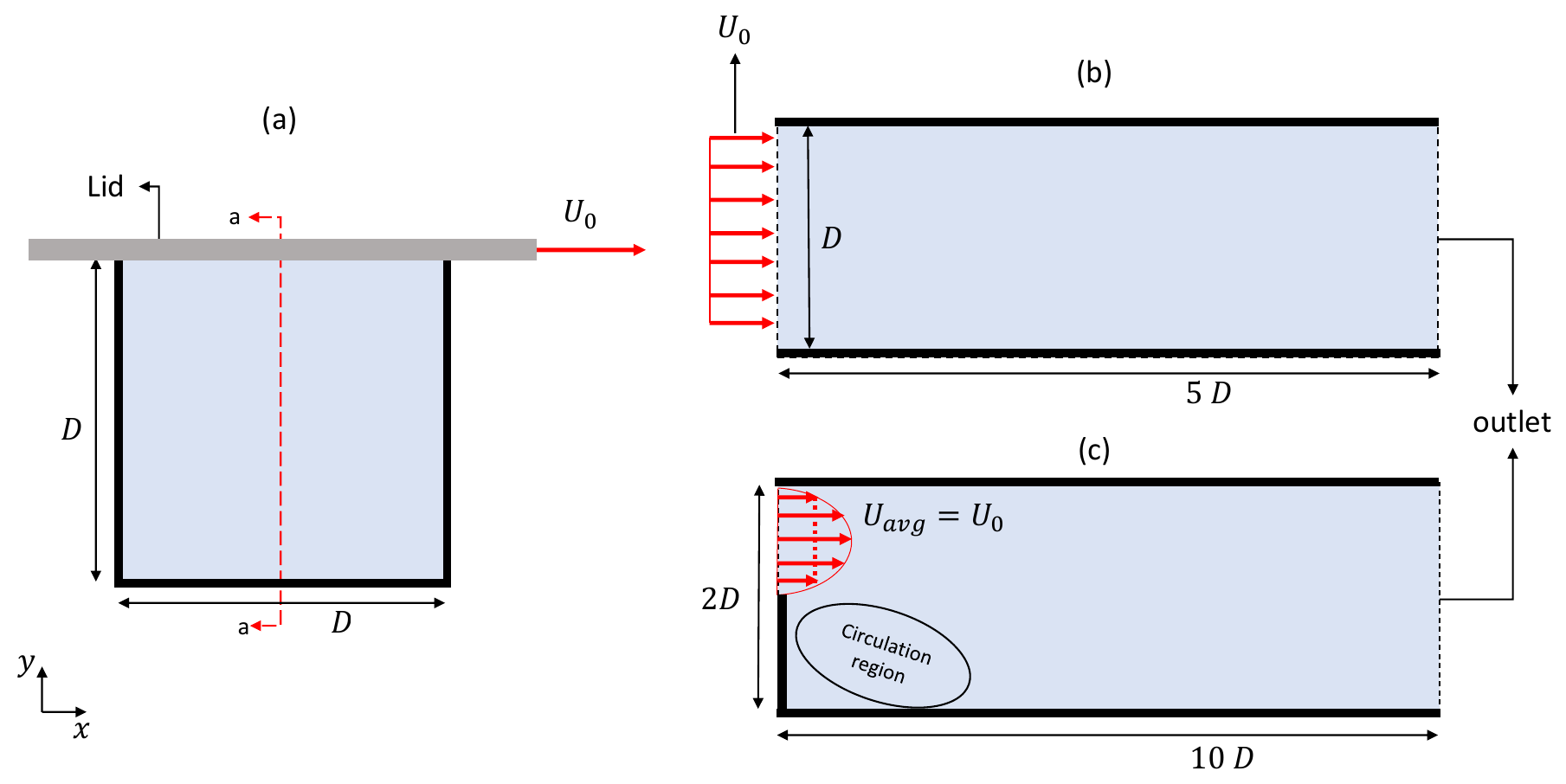}
\caption{Schematic diagram of (a) the two-dimensional lid-driven cavity problem, (b) the Poiseuille flow, and (c) the flow past a backward facing step.}\label{schematic}.
\end{figure}

To demonstrate the effectiveness of the proposed approach, we consider three benchmark examples in fluid mechanics: $i)$ the unsteady flow inside a lid-driven cavity \citep{koseff1984lid}, $ii)$ the Poiseuille flow (the unsteady laminar flow in a channel of uniform cross-section), and $iii)$ the unsteady flow in a backward facing step \citep{biswas2004backward}. As illustrated in Fig.~\ref{schematic}, the characteristic diameter, $D$, is normalized to 1, the inlet velocity, $U_0$, is set to 1, and the kinematic viscosity, $\nu$, is chosen such that the Reynolds number is 100 for all examples.  

The optimization can be carried utilizing well-established quadratic programming. Here we use MATLAB's quadratic programming toolbox, employing the \texttt{interior-point-convex} algorithm. This algorithm is tailored for large-scale optimization problems, utilizing an advanced sparse linear solver to efficiently address the substantial computational demands. The reliance on sparse solvers is particularly beneficial for problems of this nature, where the number of variables is high, but the matrices $H$ and $A$ exhibit significant sparsity \citep{nocedal2006penalty}. 

Alternatively, the vector $\mathbf{w}^*$ which minimizes Equation \eqref{eq:obj} subject to Equation \eqref{eq:bc} is given by: 
\begin{equation}
\begin{bmatrix}
\mathbf{w}^* \\
\boldsymbol{\lambda}^*
\end{bmatrix}
=
\begin{bmatrix}
H & A^T \\
A & 0
\end{bmatrix}^{-1}
\begin{bmatrix}
-\mathbf{f} \\
0
\end{bmatrix},
\label{eq:kkt}
\end{equation}
where $H$ is positive definite\footnote{For an equispaced grid similar to the one considered in this analysis, $H$ turns out to be the identity matrix and is therefore guaranteed to be positive definite.}, and $\boldsymbol{\lambda}^*$ is the vector of Lagrange multipliers. Equation \eqref{eq:kkt} is a consequence of the general result of the first-order optimality condition commonly referred to as the Karush-Kuhn-Tucker (KKT) condition \citep{nocedal2006quadratic}. 

\begin{figure}
\centering
\includegraphics[width=0.85\textwidth]{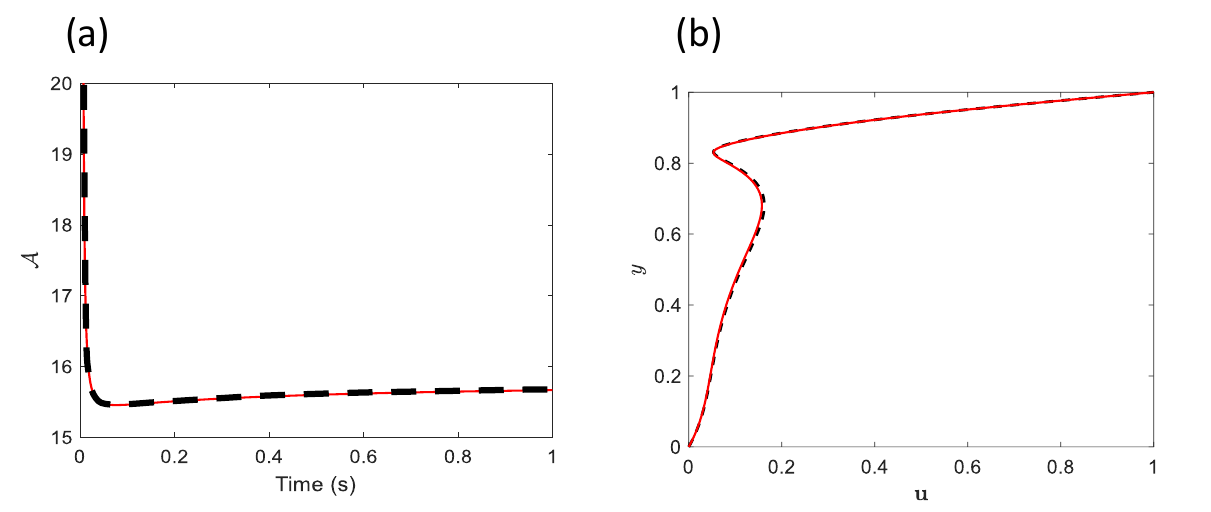}
\caption{(a) Time evolution of the minimized cost $\mathcal{A}$. (b) The flow field $\mathbf{u}$ at $t = 1$ s obtained at the a-a cross section. (Solid line) proposed minimization approach, (dashed lines) traditional high-fidelity CFD model.}\label{fig2}.
\end{figure}

Simulations for the lid-driven cavity were performed over the time range \( 0 \leq t \leq 1 \, \text{s} \). The minimized cost objective \(\mathcal{A}\) was compared with the values obtained from a traditional high-fidelity CFD model (see ref.~\citep{sababha2024variational}), revealing excellent agreement between the two approaches. As shown in Fig.~\ref{fig2}(a), \(\mathcal{A}\) initially decreases sharply, then increases at a diminishing rate, eventually stabilizing to indicate the steady-state condition, where the local acceleration approaches zero. To further validate the proposed method, we compared the predicted flow field $\mathbf{u}$ at \( t = 1 \, \text{s} \) focusing on the middle cross-section, namely (a-a) as highlighted in Fig.~\ref{schematic}. A comparison between the values of \(\mathbf{u}\) as obtained using the proposed optimization framework (solid line) and those obtained using the high-fidelity model (dashed lines) demonstrates excellent agreement as shown in Fig.~\ref{fig2}(b). 

\begin{figure}
\centering
\includegraphics[width=0.85\textwidth]{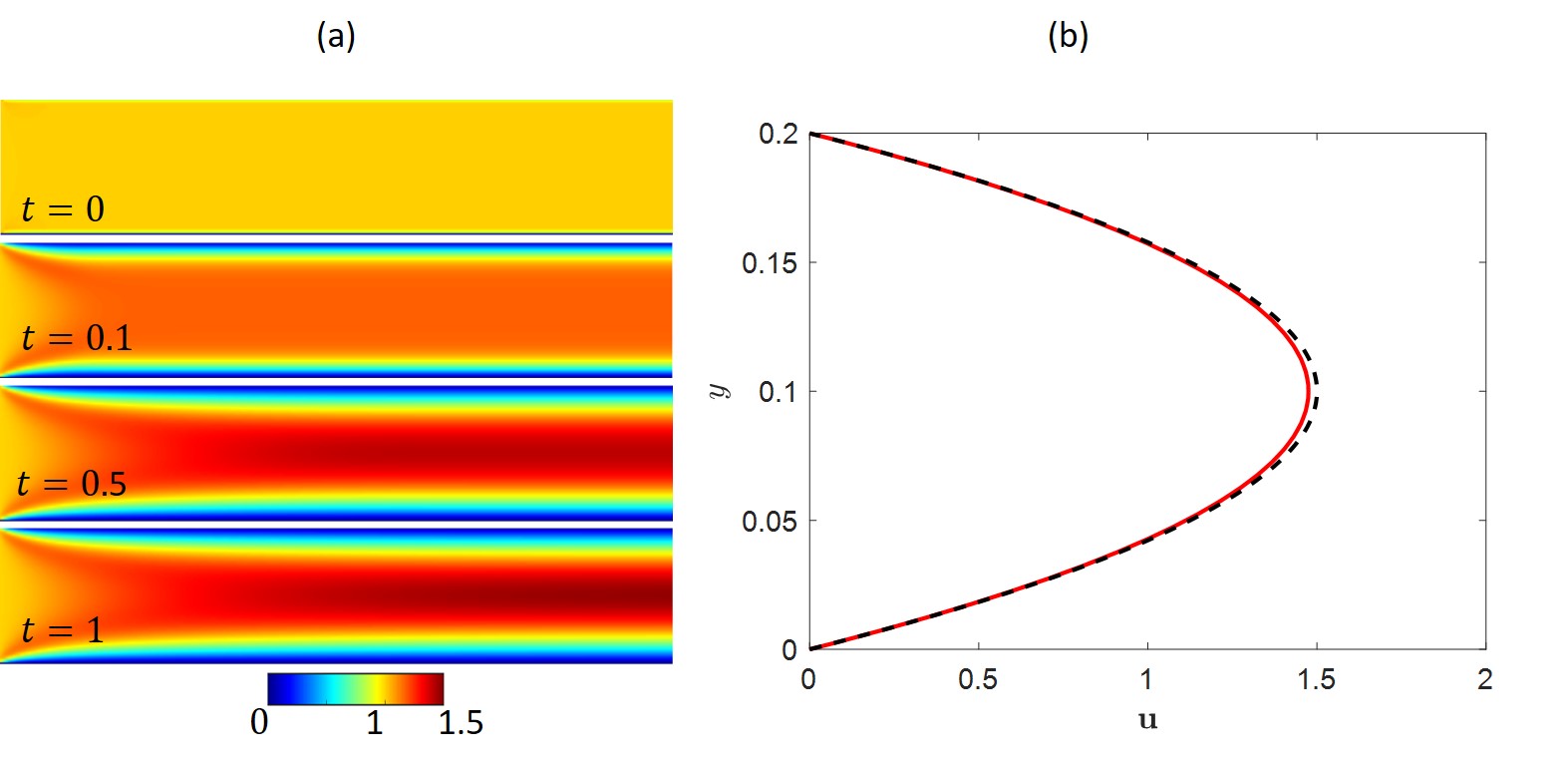}
\caption{(a) Flow field $\mathbf{u}$ at different timestamps obtained using the proposed minimization approach. (b) The predicted flow field at the outlet. (Solid line) proposed approach, (dashed lines) analytical solution.}\label{fig4}.
\end{figure}

Results for the 2D Poiseuille flow are presented in Fig.\ref{fig4}. Simulations were conducted over the time range \( 0 \leq t \leq 1 \, \text{s} \), showcasing the evolution of the developing flow at various time intervals. The transition from an initially uniform profile to a fully developed state is clearly observed, with boundary layer growth along the walls. Additionally, the predicted flow field at the outlet is shown, demonstrating a parabolic profile that aligns closely with the analytical solution derived from the Navier-Stokes equations~\citep{white2003fluid}.

\begin{figure}
\centering
\includegraphics[width=0.75\textwidth]{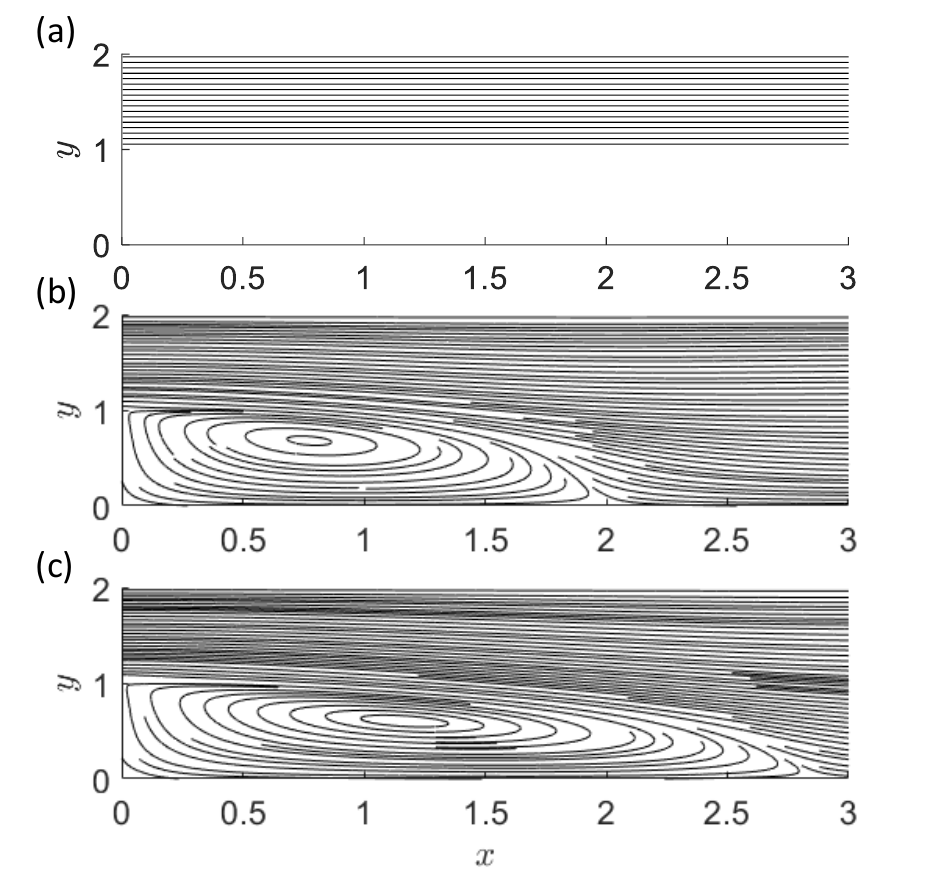}
\caption{Streamlines of the flow in the vicinity of backward step as obtained using the proposed minimization framework at $\text{Re} = 100$. (a) $t = 0$, (b) $t = 2.5$, and (c)$t = 5$.}\label{bfs}.
\end{figure}

Finally, the results for the backward-facing step are presented, with simulations conducted over the time interval \( 0 \leq t \leq 5 \, \text{s} \). The temporal evolution of the streamlines is illustrated in Fig.~\ref{bfs}, showcasing the transition from an initially uniform velocity profile to a fully developed flow state. The extent of the primary recirculation region is clearly evident, and its measured length is in excellent agreement with the numerical results reported in reference~\citep{biswas2004backward}.

\section{Concluding Remarks}
This study presents a new approach to solving incompressible fluid mechanics problems. In particular, using PMPG, we transform the CFD problem into a convex optimization framework subject to linear constraints and solve it using well-established quadratic programming tools. A direct solution can also be obtained using the well-known Karush-Kuhn-Tucker condition. The approach is implemented using the well-known unsteady lid-driven cavity problem requiring only a few lines of MATLAB code (see Appendix B), in contrast to the lengthy and complex codes often required by conventional finite volume methods, such as those used in Ansys or OpenFOAM. The results are compared to high-fidelity CFD simulations, demonstrating excellent agreement and showcasing the potential of this framework as an efficient and robust alternative to traditional CFD tools. The simplicity of the proposed method not only makes modeling fluid mechanics more accessible but also expands the potential user base, eliminating the need for custom computational frameworks.

In addition, using PMPG to cast CFD problems into a convex optimization framework offers two key advantages. First, by eliminating the pressure from the formulation, the need for pressure-velocity coupling is removed, thus avoiding the computationally expensive step of solving the Poisson equation after every time step. Second, the convexity of the problem guarantees a solution in finite time. 

Compared to the previous work of the authors~\citep{sababha2024variational}, which employed an optimization framework based on physics-informed neural networks (PINNs), this approach offers distinct advantages. First, it preserves the convexity of the problem, avoiding the challenges of non-convex training procedures in PINNs that often lead to sub-optimal performance due to the presence of local minima. Second, it ensures accurate enforcement of constraints, which are critical for obtaining physically valid solutions. PINN-based methods frequently exhibit constraint violations in certain regions, resulting in approximate but not exact solutions.

Despite these advantages, this formulation has certain limitations, as it is currently limited to square domains. Extending the method to handle irregular domains and/or domains with unstructured collocation points is a critical next step in broadening its utility. Moreover, although the advantages of the method are evident, a rigorous comparison with traditional solvers, such as finite-volume methods, is essential to fully assess the performance and applicability of the method. Benchmarking against these established tools will provide valuable information on the accuracy, computational cost, and scalability of the solution, enabling a clearer understanding of the advantages and limitations of this approach. Additionally, exploring the performance of different optimization algorithms within our framework, such as active set methods or gradient projection methods, could uncover further potential that is yet to be fully explored.

Finally, we believe that this work represents a notable step forward in addressing CFD problems. It illustrates that minimizing the pressure gradient transforms them into a convex optimization framework with guaranteed global minima that can be tackled efficiently using a wide array of well-established robust optimization tools. Furthermore, it appears that this new formulation provides a more natural approach to solving incompressible fluid mechanics. For instance, removing pressure from the formulation inherently eliminates several challenges commonly encountered in computational fluid mechanics. These include the need to solve the pressure Poisson equation~\citep{chorin1997numerical, toutant2018numerical}, the complexity of prescribing appropriate boundary conditions~\citep{papanastasiou1992new}, and issues such as spurious oscillations that arise from interpolation in collocated grids~\citep{zhang2014generalized}. Nevertheless, while pressure is removed from the formulation, it can still be recovered from the Lagrange multipliers in Equation \eqref{eq:kkt}, since the pressure serves as a Lagrange multiplier enforcing the continuity constraint~\citep{gresho1987pressure}.

\appendix
\newpage
\section{Code Implementation}

The MATLAB code for solving the lid-driven cavity problem using quadratic programming is presented below. It is evident that this approach is more concise and straightforward compared to traditional methods. 



\begin{lstlisting}[language=Matlab, caption={MATLAB code for solving the optimization problem}, label={lst:optimization}, basicstyle=\footnotesize\ttfamily, breaklines=true, frame=single]
%% 
clc; clear; close all
%% --- define geometry and boundaries ----
n = 201; 
tf = 0.5; 
dt = 0.001; 
lx = 1; 
ly = 1; 
nsteps = 2; 
Re = 1e2; 
n_vars = n^2*2;
%------------------------------
nt = ceil(tf/dt); 
dt = tf/nt;
x = linspace(0,lx,n); 
y = linspace(0,ly,n); 
[X,Y] = meshgrid(y,x);
h = lx/(n-1); 
%-----------------------------
nu = 1/Re; 
U = zeros(n); V = zeros(n); 
U(n, :) = 1;
x0 = [];

H = 2*speye(n_vars); 
m = size(A, 1); 
KKT = [H, Aeq'; Aeq, sparse(m, m)];
for i = 1:nt
%% --------pre-processing --------
disp(['i = ' num2str(i)])
[Ux, Uy] = gradient(U, h); [Vx, Vy] = gradient(V, h);
[Uxx, ~] = gradient(Ux, h); [Vxx, ~] = gradient(Vx, h);
[~, Uyy] = gradient(Uy, h); [~, Vyy] = gradient(Vy, h);
a = U.*Ux + V.*Uy - nu.*(Uxx + Uyy);
b = U.*Vx + V.*Vy - nu.*(Vxx + Vyy);
a = a'; b = b';
f = 2.*[a(:); b(:)];

%% --------Solution through Quadratic Programming --------
options = optimoptions('quadprog','Display', 'none', 'Algorithm', 'interior-point-convex');
 %'TolFun', 1e-2, 'TolCon', 1e-2
[optimal_Ut, fval, exitflag, output, lambda]  = quadprog(H, f, [], [], Aeq, beq, [], [], x0, options);

%% --------Solution through Karush-Kuhn-Tucker Condition----
B = [-f; sparse(m, 1)]; 
optimal_Ut = KKT\B;

%% --------postprocessing and March in time --------
ut = optimal_Ut(1: n^2); vt = optimal_Ut(n^2 + 1: end);
Ut = reshape(ut, n, n)';
Vt = reshape(vt, n, n)';
U = U + dt.* Ut;
V = V + dt.* Vt;
end

\end{lstlisting}
\newpage
\bibliographystyle{jfm}
\bibliography{jfm}
\end{document}